\documentclass[journal, 12pt]{IEEEtran}
\usepackage{latexsym}
\usepackage{epsfig}
\usepackage{amsfonts,amsmath,amssymb,amstext}
\usepackage{epstopdf}
\usepackage{graphicx}
\usepackage{psfrag}
\usepackage{dsfont}
\usepackage{color}
\usepackage{cite}
\usepackage{float}

\begin{document}
	\title{Cooperative Hybrid Networks with Active \\ Relays and RISs for B5G: Applications, \\ Challenges, and Research Directions}

\author{Zaid~Abdullah,
	Steven~Kisseleff, Wallace~Alves~Martins, Gaojie~Chen, 
	
	Luca~Sanguinetti, Konstantinos~Ntontin, Anastasios~Papazafeiropoulos, Symeon~Chatzinotas, and~Bj$\ddot{\text{o}}$rn~Ottersten%
	\thanks{This work has been submitted to the IEEE for possible publication. Copyright may be transferred without notice, after which this version may no longer be accessible.\\ This work was supported by the Luxembourg National Research Fund (FNR) through the CORE Project under Grant RISOTTI C20/IS/14773976.
		\par Z. Abdullah, S. Kisseleff, W. A. Martins, K. Ntontin, S. Chatzinotas, and B. Ottersten are with the SnT centre, University of Luxembourg, L-1855, Luxembourg.
	\par G. Chen is with the Institute for Communication Systems (ICS), 5GIC \& 6GIC, University of Surrey, Guildford, GU2 7XH, U.K. 
		\par A. Papazafeiropoulos is with the Communications and Intelligent	Systems Research Group, University of Hertfordshire, Hatfield AL10 9AB, U.K.
			\par L. Sanguinetti is with the University of Pisa, Dipartimento di Ingegneria dell’Informazione, 56122 Pisa, Italy. 
			
			\par \textit{Corresponding authors: Zaid Abdullah (zaid.abdullah@uni.lu) and Gaojie Chen (gaojie.chen@surrey.ac.uk).}}}

\maketitle

\begin{abstract}
Among the recent advances and innovations in wireless technologies, reconfigurable intelligent surfaces (RISs) have received much attention and are envisioned to be one of the enabling technologies for beyond 5G (B5G) networks. On the other hand, active (or classical) cooperative relays have played a key role in providing reliable and power-efficient communications in previous wireless generations. In this article, we focus on hybrid network architectures that amalgamate both active relays and RISs. The operation concept and protocols of each technology are first discussed. Subsequently, we present multiple use cases of cooperative hybrid networks where both active relays and RISs can coexist harmoniously for enhanced rate performance. Furthermore, a case study is provided which demonstrates the achievable rate performance of a communication network assisted by either an active relay, an RIS, or both, and with different relaying protocols. Finally, we provide the reader with the challenges and key research directions in this area. 
\end{abstract}

\section{Introduction}\label{introduction}
In this section, we start by discussing the concepts and protocols of active relaying and reconfigurable intelligent surfaces (RISs). Subsequently, we introduce and motivate the deployment of hybrid relaying schemes.  
\subsection{Active Relaying: Concept and Protocols}
The utilization of cooperative devices with active relaying capabilities can lead to a substantial improvement in network throughput \cite{laneman2004cooperative}, and over the last decades, more and more applications involving various types of relays have been proposed and implemented \cite{relayStand}. 
The simplest concept of a relay is the reception, amplification, and re-transmission of a signal. The corresponding protocol is known as amplify-and-forward (AF). In this context, it is important to mention that the AF protocol unavoidably leads to noise amplification, which is especially harmful if the received signal at the relay is of low quality.
\par Another prominent relaying protocol is the decode-and-forward (DF), where the received signal undergoes a complete demodulation and decoding procedure as well as re-encoding and re-modulation, and thereby avoiding the noise amplification drawback of the AF protocol. The DF strategy enables the split of the communication link into two independent sub-links, such that the overall throughput corresponds to the lowest throughput among the two sub-links. Despite the advantages of the DF relays in terms of achievable throughput, they suffer from higher delays and complexity due to the required signal processing operations.
\par Although there are many other relaying protocols (e.g. Filter-and-forward, quantize-and-forward, context-aware, and buffer-aided relays), in this article we focus on the AF and DF relays given their widespread deployment in modern telecommunications.  
\par It is worth highlighting that in case of identical channel characteristics for both sub-links of a relay-assisted communication link with properly optimized output powers, it can be observed that the attainable throughput is maximal, if the relay is placed exactly in the middle between the transmitter and the receiver. Interestingly, this observation is common for most types of active relays over wireless channels. For convenience, throughout this article we adopt the term \textit{relay} when referring to an active relay device.

\subsection{RISs: Concept, Architecture and Operation}
RISs have recently emerged as an attractive solution to meet the ever-increasing demand of higher data rates over wireless channels, while maintaining low-cost and low-power requirements \cite{pan2021reconfigurable}. Specifically, an RIS can tweak the (often unpredictable) response of a wireless channel by means of smart reflections of impinging signals on its planar surface. 

\par RISs can be implemented in various ways, including \textit{reflect-arrays} and \textit{software-defined meta-materials} \cite{basar2019wireless}. Each surface consists of a large number of digitally controlled, small meta-atoms, also known as unit cells (UCs). Unlike  multi-antenna relays, RISs are meant to be energy-efficient nodes, and hence, they only perform nearly-passive controlled signal reflections of impinging electromagnetic waves with certain phase and/or amplitude adjustments.\footnote{For the similarities and differences between relays and RISs, we refer the reader to the work in \cite{di2020reconfigurable} and the references therein.} With such limited processing capabilities, the response of each UC at the RIS, also known as \textit{passive beamforming}, is often optimized at an external node and then communicated to a control unit that is connected to the RIS \cite{wu2019towards}. \par It is worth highlighting that unlike the relay case, the highest signal power enhancement provided by an RIS occurs when the RIS is closest to the transmitter or the receiver, but not in the middle. This is the result of the double path-loss of the overall channel gain of RIS-assisted networks, i.e. the multiplication of channel gains of first and second communication sub-links, which can be a performance-limiting factor given the absence of active power amplification capabilities at the RIS. 
\par Recently, RISs have received an unprecedented attention from researchers around the globe, and they are envisioned to have a key role in future wireless networks with a plethora of practical applications \cite{pan2021reconfigurable, basar2019wireless, di2020reconfigurable, wu2019towards}. 

\begin{figure}[t]
\centering
\includegraphics[scale=1]{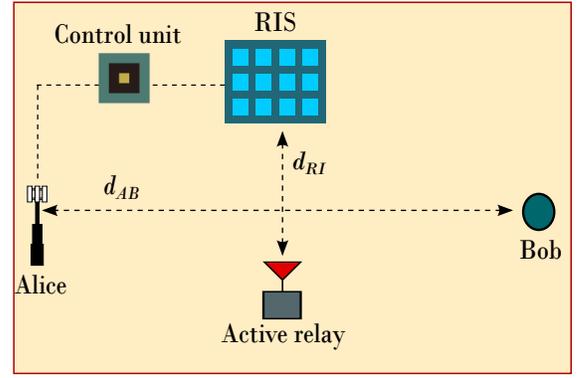}
\caption{A hybrid relaying network with a relay and an RIS. The distances between Alice to Bob, and relay to RIS are denoted by $d_{AB}$ and $d_{RI}$, respectively.  }
\label{figure1}
\end{figure}
\subsection{Hybrid Relaying Architectures}\label{hybrid}
In hybrid relaying networks (HRNs), both relays and RISs are utilized simultaneously to assist the communication between two transceiving nodes \cite{abdullah2020hybrid} (see Fig.~\ref{figure1}). The benefits of such HRNs can be highlighted by answering the following two questions: (Q1) \textit{`how can an RIS improve the performance of a relay-assisted network?'}, and (Q2) \textit{`how can a relay improve the performance of an RIS-assisted network?'}. 
\par Before answering the two questions above, we first need to clarify that throughout this article, the term \textit{relay-assisted} refers to the case where only a relay is employed to assist the communication between two transceiving nodes, while the term \textit{RIS-assisted} refers to the case where only an RIS is utilized to facilitate the communication. For convenience, we will contemplate on a one-way relaying scenario to answer the above two questions, while the benefits of other HRNs will be discussed in Section \ref{Use cases}. 

\par Let Alice and Bob be the transmitter and receiver nodes, respectively, and assume that there are one relay and one RIS placed between Alice and Bob, as depicted in Fig.~ \ref{figure1}. Furthermore, assume that there are direct links between all different nodes except between Alice and Bob due to signal blockage for instance. Thus, communication can only be facilitated via incorporating either the relay, the RIS, or both. 
\par Regarding the answer to question (Q1), \textit{the RIS can always enhance the rate performance of a relay-assisted network, as long as its phase-shifts are properly set} \cite{abdullah2020hybrid}. Specifically, the RIS can increase the spatial diversity of the two communication links, i.e. the link between Alice and the relay, and that between the relay and Bob. In addition, the RIS can also be utilized to maximize the minimum received signal-to-interference-plus-noise ratio (SINR) at the relay and Bob when the relay operates in a full-duplex (FD) mode \cite{abdullah2020optimization}. However, the extent of the performance improvement that an RIS can provide to a relay-assisted network depends on different factors that will be discussed in the next section. 
\par In contrast, and regarding the answer to question (Q2), \textit{the main benefit of deploying a relay to support an RIS-assisted network is to reduce the double path-loss effects}. Specifically, placing a relay near the RIS dramatically enhances the channel conditions, especially if the RIS was far away from both Alice and Bob. However, as RISs operate in an FD fashion (i.e. they do not introduce large signal delays), utilizing an half-duplex (HD) relay might not always improve the rate performance due to the limited bandwidth as will be discussed in the following section. 
\par It is worth highlighting that, although the recently proposed active RISs can potentially minimize the double path loss by amplifying the impinging signals on their surfaces \cite{long2021active}, relays bring unique advantages in terms of signal processing capabilities. In addition, and unlike active RISs, relays can be utilized for various applications as cooperative nodes as will be discussed throughout this manuscript. 
\par In the reminder of the article, we first introduce various use cases of different HRNs. Subsequently, we evaluate and analyze the channel gain and spectral efficiency of different relay-assisted, RIS-assisted, and HRNs. The main challenges, research directions, and practical applications of HRNs are then presented and discussed. Concluding remarks are provided at the end of the article. 
\section{Use Cases} \label{Use cases}
In this section, we present different hybrid architectures where both the relay and the RIS contribute to the enhancement of information exchange between Alice and Bob. 
\begin{figure*}[t]
    \centering
    \includegraphics[width=17cm,height=10cm]{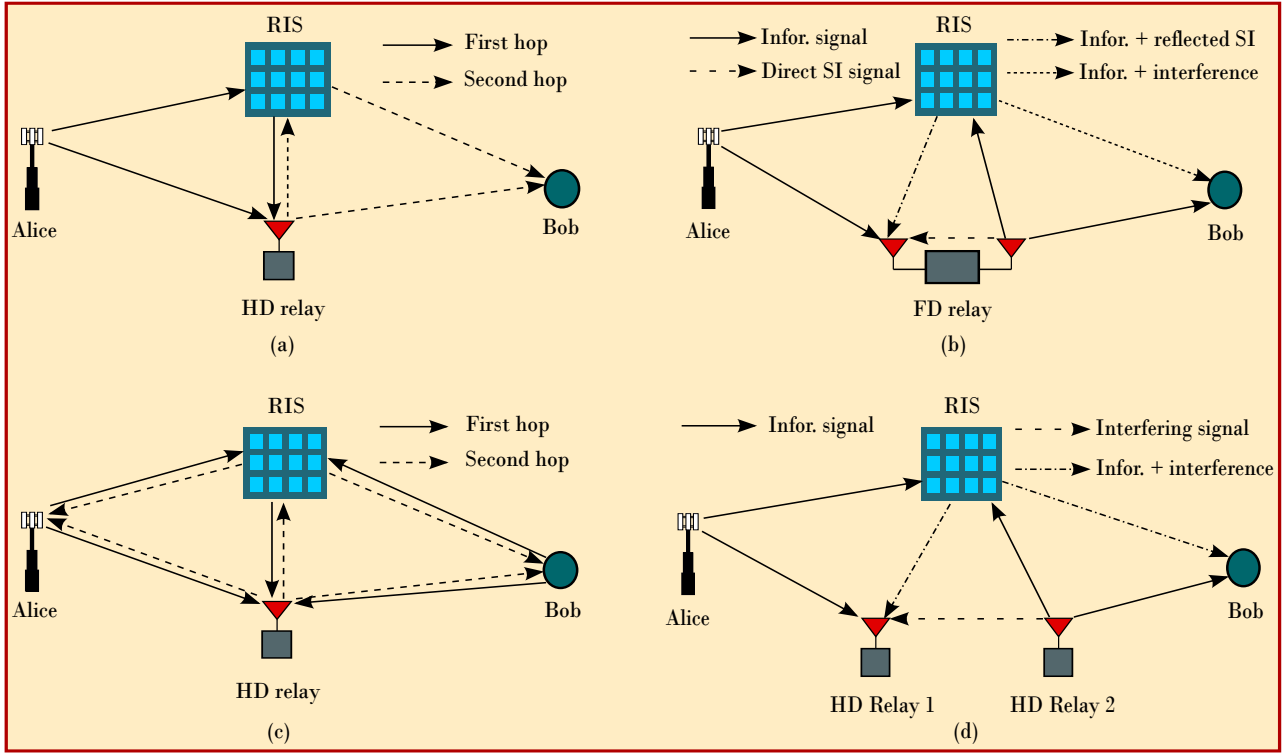}
    \caption{Use cases of HRNs: (a) Hybrid one-way relaying with an HD relay, (b) Hybrid one-way relaying with an FD relay, (c) Hybrid two-way relaying, and (d) Hybrid successive relaying when Relay 1 and Relay 2 operate in receive and transmit modes, respectively.  }
    \label{figure2}
\end{figure*}
\subsection{Hybrid One-way Relaying}
This case was briefly discussed in Section \ref{hybrid} to motivate the use of HRNs. Nonetheless, here we will further elaborate on such a hybrid network when both HD and FD relays are adopted. 

\subsubsection{HD relays}
When an HD relay is present, one deployment strategy for the RIS is to adjust its phase-shifts to maximize the received signals' quality at the relay and Bob during first and second hops, respectively (see Fig. \ref{figure2}a).\footnote{Note that there are other deployment strategies for the RIS in such hybrid networks as proposed, for example, in \cite{yildirim2021hybrid}.} The extent of improvement that an RIS can provide when supporting a relay-assisted network depends on the accuracy of phase-configuration, the number of UCs, and the gain of channels between the relay and the RIS. With a proper phase-adjustment at the RIS, the performance of a relay-assisted network can always be improved in such scenarios \cite{abdullah2020hybrid}.  
\par In contrast, the RIS operates in an FD fashion, which means that utilizing an HD relay might hinder the performance of an RIS-assisted network due to the inefficient bandwidth utilization of the HD relay. Nonetheless, it was demonstrated in \cite{abdullah2020hybrid} that unless the transmit power at Alice and number of UCs at the RIS are both very large, an HD relay can provide large performance gains to RIS-assisted networks. The reason is that when the amount of radiated power and/or number of UCs are/is limited, the network is in the power-limited regime. Thus, deploying a relay can overcome the effects of the double path-loss, thereby resulting in higher achievable rates. 
\par One can conclude from the analysis above that when operating in the power-limited regime, an HRN consisting of an HD relay and an RIS can achieve superior data rates compared to relay-assisted or RIS-assisted networks. 
\subsubsection{FD relays}
To overcome the bandwidth limitation of HD relaying, FD relays can be adopted as depicted in Fig.~\ref{figure2}b. Here, the RIS can be utilized to maximize the received signal powers at the receive antenna of the relay and Bob, transmitted from Alice and the transmit antenna of the relay, respectively. In addition, the RIS can be configured to mitigate the interfering signal from Alice to Bob, and also minimize the effects of direct and/or reflected (i.e. through the RIS) self-interference (SI) signal from the transmit to the receive antenna at the relay. \par Clearly, the rate performance of an FD relay-assisted network can be enhanced when an RIS is deployed. On the other hand, the superiority of such an HRN over an RIS-assisted network depends on both the type of relay deployed, and the level of residual SI. In particular, when efficient relaying protocols such as the DF are deployed, and the residual SI is suppressed to low levels, an HRN with FD relaying can outperform an RIS-assisted system even when both the transmit power and number of UCs at the RIS are large \cite{abdullah2020optimization}. In contrast, AF relays lead to noise and residual SI amplification, and thus, an RIS-assisted network can be superior to an HRN with an FD-AF relay especially when operating in the bandwidth-limited regime with large number of UCs, as will be demonstrated in Section \ref{evaluations}. 

\subsection{Hybrid Two-way Relaying}
When both Alice and Bob wish to exchange their data via an HD relay, one can adopt a two-way relaying protocol for efficient spectrum utilization. In particular, both Alice and Bob, who work in HD mode, transmit their information to the relay during the first-hop. During the second hop, the relay broadcasts its received signal after a proper power amplification (see Fig.~\ref{figure2}c). Here, the RIS can be deployed to provide spatial diversity and enhance the rate performance as in \cite{wang2021joint}, minimize the transmit powers of active nodes, and/or provide over-the-air cancellation of SI signals at Alice and Bob.
\subsection{Hybrid Successive Relaying}
In successive relaying (SR), two HD relays are deployed to utilize the full bandwidth by mimicking the operation of an FD relay. At any given time, one relay receives a signal from Alice, while the other relay transmits the received signal from Alice in the previous time instant to Bob (see Fig.~\ref{figure2}d). Thus, a key challenge in SR is the resultant inter-relay interference (IRI). To that end, RISs can be utilized to mitigate the effects of IRI while maximizng the power of useful signals \cite{successive}.   
\section{Evaluations} \label{evaluations}
We consider a scenario where each active node (Alice, relay, and Bob) has a single radiating element, while the RIS contains $M$ UCs. The antenna gain at Alice and Bob is $0$ $\mathrm{dBi}$, while a $5$ $\mathrm{dBi}$ antenna gain is assumed for the relay and RIS. A 2D network setup is considered similar to that shown in Fig.~\ref{figure1}, where Alice is located at $(0,0)$, while Bob is located at $(d_{AB}, 0)$ with $d_{AB}$ being the distance between Alice and Bob in meters. In addition, we adopt a symmetric HRN where the locations of relay and RIS are set as $(\frac{d_{AB}}{2}, -\frac{d_{RI}}{2})$ and $(\frac{d_{AB}}{2}, \frac{d_{RI}}{2})$ with $d_{RI}$ being the distance between relay and RIS. Direct links exist between all nodes except from Alice to Bob. The carrier frequency is $3$~GHz, and the noise power is $\sigma^2= -94$ $\mathrm{dBm}$. Perfect channel state information (CSI) is assumed to be available with an ideal reflection amplitude of $1$ per UC. The total transmit power at any transmission time is $P_T$ Watts, and thus when an FD-relay is involved, both Alice and the relay transmit with power level of $0.5P_T$ Watts. 
\begin{figure}[t]
    \centering
    \includegraphics[scale=0.65]{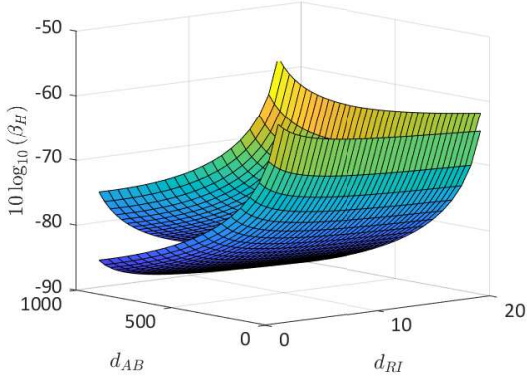} 
    \caption{Distance-based channel gain of an HRN vs. distances (in meters) between Alice to Bob, and RIS to relay when $M=64$ (bottom surface), and $M=400$ (top surface).}
    \label{fig3}
\end{figure}

\subsection{Channel Gain Analysis}
\par Fig.~\ref{fig3} illustrates the channel gain of the HRN ($\beta_H$) as a function of $d_{AB}$ and $d_{RI}$ over free space propagation and perfect phase-adjustment at the RIS. As one would expect, when $d_{AB}$ and $d_{RI}$ increase, the channel conditions become more challenging due to increased overall path-loss. However, larger surfaces with a higher number of UCs lead to improved channel conditions. Note that $\beta_H$ represents the overall channel gain of one hop, and since the considered HRN is symmetric, the effective channel gains are identical for both first and second hops.

\begin{figure}[t]
    \centering
    \includegraphics[scale=0.65]{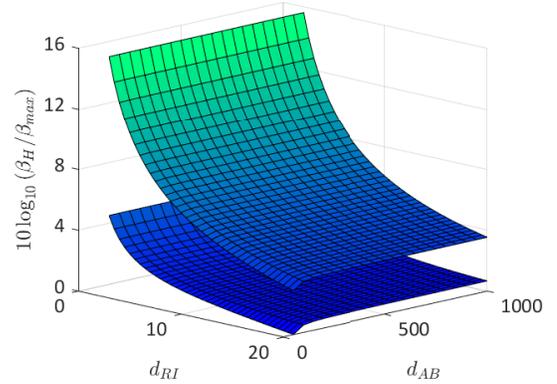}
    \caption{Channel gain improvement of an HRN over non-hybrid relaying schemes as a function of distances (in meters) between Alice to Bob, and RIS to relay (for the HRN) when $M = 64$ (bottom surface), and $M = 400$ (top surface). The parameter $\beta_{max}$ represents the maximum channel gain between the RIS-assisted and relay-assisted cases. }
    \label{fig4}
\end{figure}
\par The superiority of HRNs over both RIS-assisted and relay-assisted schemes can be seen in Fig. \ref{fig4}. Particularly, we show the improvement in channel gain over free space propagation when a symmetric HRN is utilized compared to the case where only either an RIS or a relay is deployed. For the RIS-assisted case, the RIS placement was fixed at a close proximity to Alice at $(0,10)$ meters to minimize the double path-loss effect. For the relay-assisted case, the relay was placed exactly between Alice and Bob at $(\frac{d_{AB}}{2},0)$ meters. Interestingly, the distance between Alice and Bob has a negligible impact on the comparison between hybrid relaying and non-hybrid (i.e. relay-assisted and/or RIS-assisted) networks. However, the distance between the relay and RIS for the HRN plays a key role in the comparison, as the closer the two nodes are to each other, the higher the gain becomes for the HRN over non-hybrid schemes. Nonetheless, when $d_{RI} = 10$ meters, which is the same distance between Alice and RIS for the RIS-assisted case, the HRN demonstrates an impressive $6$ $\mathrm{dB}$ improvement in channel gain compared to the non-hybrid relaying, given that the RIS is equipped with $400$ UCs.
 \begin{figure}[t]
    \centering
    \includegraphics[scale=0.65]{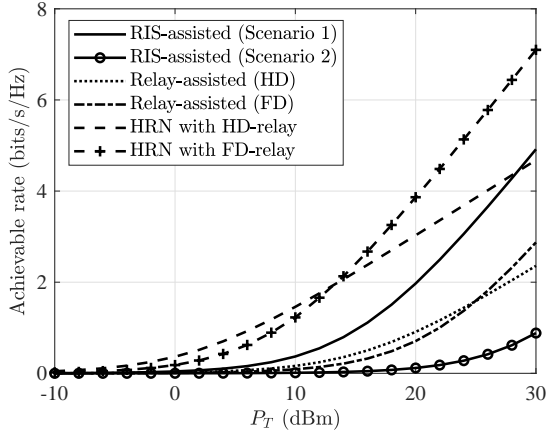}
    \caption{Achievable rate vs. transmit power for different relaying schemes when $M = 64$, and $\gamma_{SI}/\sigma^2 = 0$ dB. The DF protocol was adopted for all cases that include active relaying (i.e. HRNs and relay-assisted cases).}
    \label{fig5}
\end{figure}
\subsection{Spectral Efficiency Analysis}
\par
\begin{figure}[t]
    \centering
    \includegraphics[scale=0.65]{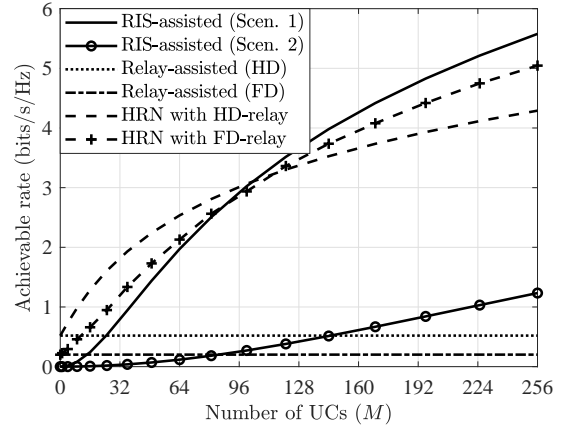}
    \caption{Achievable rate vs. number of UCs at RIS ($M$) for different relaying schemes when $P_T = 20$ dBm, and $\gamma_{SI}/\sigma^2 = 5$ dB. The AF protocol was adopted for all cases where active relaying is involved. }
    \label{fig6}
\end{figure}

 In Fig.~\ref{fig5} and Fig.~\ref{fig6}, we adopt the 3GPP Urban Micro (UMi) channel gain model \cite {3GPP}. We neglect the shadow fading effects and study the achievable rate performance of one-way relaying over deterministic flat-fading channels. Specifically, given that RISs can be installed on tall buildings, all links from/to the RIS were modeled as line-of-sight (LoS). In contrast, relays can be cooperative mobile users in a dense urban network, and hence, links between the relay and both Alice and Bob were modeled as non-LoS. A symmetric HRN is adopted with $d_{RI} = 15$~m and $d_{AB} = 300$~m. For the RIS-assisted case, we consider two scenarios to demonstrate the effect of the double path loss on the rate performance. \textbf{Scenario~1:} The RIS is located near Alice at $(0, d_{RI})$, and \textbf{Scenario~2:} The RIS is located between Alice and Bob at $(\frac{d_{AB}}{2}, \frac{d_{RI}}{2})$. For the relay-assisted network, the relay was located in the middle at $(\frac{d_{AB}}{2},0)$. In addition, for the HRN with FD relays, deterministic channels are generated with arbitrary phases, and the particle swarm optimization method in \cite{successive} was deployed to optimize the rate with 500 particles and 100 optimization iterations under an auxiliary convergence parameter of $\pi/8$ \cite{successive}. Furthermore, the residual SI ($\gamma_{SI}$) includes both the direct and the reflected loop interference for the HRNs with FD-relays.  

\par Fig. \ref{fig5} illustrates the achievable rate performance as a function of transmit power levels. Notably, the RIS-assisted case in Scenario~2 shows the worst performance among all other schemes, which demonstrates the significant impact of the double path loss when the RIS is far away from both the transmitter and the receiver. In contrast, the HRN with FD relay achieves the highest rate at high transmit power regime, including the RIS-assisted case with an optimally placed RIS (i.e. Scenario~1), given that the SI is suppressed to the noise level. This demonstrates that utilizing an efficient FD-DF relay can enhance the performance of RIS-assisted networks even at high levels of transmit powers, i.e. when the network is operating in the bandwidth-limited regime. Moreover, the HRN with an HD relay shows superior performance compared to the RIS-assisted case at low and medium levels of transmit powers, i.e. when the network operates in the power-limited regime. However, at high levels of transmit powers, an RIS-assisted case with a minimal double path-loss becomes superior due to the bandwidth limitation of the HD relaying. Furthermore, the achievable rates for HRN with HD/FD relaying are always higher than the HD/FD relay-assisted cases, which shows that RISs can indeed always enhance the performance of relay-assisted networks.

\par Finally, Fig. \ref{fig6} demonstrates the rate performance of different relaying network architectures for a wide range of values of UCs at the RIS. The results show that under AF relaying with a non-negligible residual SI, the RIS-assisted case in Scenario~1 can outperform even the HRN with FD relay when the number of UCs is relatively large. This is due to the fact that the AF relay introduces noise and residual SI amplifications when operating in the FD mode, which can be costly when dealing with high levels of residual SI. Nonetheless, such HRNs with HD/FD AF relays are still preferable over the RIS-assisted case, if the RIS was far away from both the transmitter and receiver.

From the analysis above, one can conclude that utilizing a cooperative relaying device, regardless of its type and operational mode, can bring large performance improvements to an RIS-assisted transmission when the RIS is far away from both the transmitter and the receiver. Moreover, when the network is in the power limited regime, an HRN is preferable over an RIS-aided transmission even if the RIS is optimally placed. However, to challenge the performance of an optimally placed RIS in the bandwidth limited regime, one would require an HRN with efficient FD relaying protocol and a negligible amount of residual SI.

\section{Research directions and applications}
Here, we provide the key challenges and research directions for HRNs from a practical standpoint. Then, we introduce various applications for such HRNs.
\subsection{Future Research Directions}
\subsubsection{CSI acquisition}
The performance enhancement achieved by incorporating an RIS into any communication network highly depends on the efficiency (or accuracy) of the passive beamforming design of the RIS. This entails having some knowledge about the channel models corresponding to the nodes sending/receiving signals to/from the RIS.

In general, the channel estimation (CE) task in an RIS-aided transmission is performed at the transmitter side or the receiver side, but not at the RIS due to its nearly passive nature. In addition, links between two tranceiving nodes corresponding to different UCs at the RIS (or corresponding to the same UC but for different active nodes) need to be estimated separately. As such, the amount of training required for CE depends on the size of the RIS as well as the number of active nodes in the network. This makes the CSI acquisition a key challenge \cite{wei2021channel}, which might be made even more difficult in HRNs with the deployment of relaying devices.

Specifically, and compared to an HD point-to-point RIS-aided transmission, the amount of training required in HRNs can increase notably when more than two transceiving nodes are active at the same time. Such a case appears, for example, in an HRN utilizing an FD relay. Nonetheless, as the use of relays tends to boost the performance of RIS-assisted networks (except in the cases already detailed), one should account for the CE overheads when designing the whole hybrid network, including not only the definition of number, type, and operational mode (i.e. HD vs FD) of relays, but also the number of UCs at the RIS, which in turn has a direct impact upon the complexity of the CSI acquisition task. Therefore, investigating the performance of HRNs while considering the required overheads and/or designing novel CE schemes for HRNs is a key research direction.

\subsubsection{Centralized vs distributed processing}
In centralized implementations, a central entity controls the communication-related activities of the other network entities, including interference management, CSI acquisition, beamforming design, etc. This architecture is particularly useful for interference management, but might actually be infeasible in a fast-changing wireless environment especially when dealing with large scale HRNs utilizing many RISs and relays, given the huge computational burden and bandwidth requirements over control channels. However, centralized processing might still be a feasible solution in such HRNs if the optimization of RISs was carried out based on the statistics of the CSI, which vary very slowly in practice and thus can significantly reduce the frequency of phase-reconfiguration at the cost of a degraded performance.  

\par Alternatively, one can adopt a distributed processing architecture, where some sensing and processing functionalities are required for the network nodes to be able to autonomously optimize themselves based on the environmental state information and network configuration. We expect that this burden would be on the relays, which would then work in coordination with the control-unit of the RIS to implement distributed optimization algorithms. Such an approach relies on the signal processing capabilities of the relays, and would leverage on fully distributed estimation/optimization approaches that try to keep at minimum the amount of signaling across different nodes to achieve consensus over the estimated/optimized parameters. In this context, the inherent signaling overhead of a centralized solution would be traded off by an increased computational burden over the relays.

\subsubsection{Relay/RIS deployment and route optimization}
The gain of HRNs over RIS-assisted networks mainly comes from minimizing the double path-loss of the latter case. This means that for designing efficient HRNs, efficient relay/RIS pairing can play a key role. Therefore, one can think of relay-placement and/or relay-selection as possible ways to minimize the double path-loss effects, or even carry out a joint relay/RIS pairing with passive beamforming design for optimal rate performance~\cite{Chen211}.

\par In addition, to extend the coverage and connectivity while minimizing the total transmit power, multi-hop communication is adopted in practice. In such scenarios, one can take advantage of the large number of available relays and RISs via route optimization schemes, which can lead to large gains in both rate and energy-efficiency performance. Furthermore, when multiple RISs are present, it is possible to design power control schemes and passive beamforming for RISs while considering signal reflections through multiple RISs and also inter-RIS reflections, leading to an enhanced end-to-end performance. In this context, tools from machine learning (such as deep reinforcement learning) can be leveraged to provide efficient solutions in such dynamic environments.

\subsubsection{Energy efficiency}
Thus far, the works on HRNs have been mainly focused on either enhancing the achievable rate or minimizing the total transmit power of a communication system \cite{abdullah2020hybrid, yildirim2021hybrid, abdullah2020optimization, wang2021joint, successive, Chen211}. Indeed, in many cases, utilizing both relays and RISs can be an attractive choice regarding those aspects. However, and unlike relay- or RIS-assisted networks, HRNs include both relays and RISs, leading to higher energy consumption. Thus, the obtained rate enhancement and/or savings in transmit powers do/does not necessarily mean better energy efficiency performance of HRNs compared to non-hybrid schemes. Therefore, an investigation into the energy efficiency performance of different HRNs compared to their counterpart is of great necessity.

\subsection{Practical Applications}
\subsubsection{Cognitive radio (CR) networks}
In CR, two groups known as \textit{primary network} (PN) and \textit{secondary network} (SN), share the same licensed spectrum for efficient spectrum utilization. However, power restrictions at the SN are applied to ensure that the interference level at the PN is below a predefined threshold. 

\par In such cases, an HRN with a relay and an RIS can provide large improvements in achievable rates. Specifically, and due to power restrictions imposed by the PN, the SN tends to operate in the power-limited regime. Thus, deploying the relay, either on its own or with the RIS, at the SN can lead to substantial improvements in data throughput of SN. On the other hand, the RIS can be deployed either near the relay, or at the PN side. In any case, the RIS can provide enhanced signal quality at PN and SN (including the relay sub-links), while performing over-the-air cancellation or mitigation of interfering signals from the primary transmitter to the relay and the SN receiver. 

\subsubsection{Unmanned aerial vehicle (UAV) networks} UAVs can act as aerial
relays and extend the coverage of terrestrial communication. They can leverage their high elevation compared with terrestrial nodes, by either mounting a relay node or an RIS. In the former case, the UAV battery could potentially be drained in a much shorter amount of time due to power-hungry active components, mainly power amplifiers. Hence, UAVs that perform relaying operations through mounted nearly-passive RISs seem a more viable approach from
an energy-consumption point of view. 
\par In hybrid scenarios, an RIS mounted on a UAV could be used for directing the signal through reflection to a terrestrially placed relay for the desirable active amplification before the signal is dispatched to the destination. Such a scenario would be highly beneficial in the uplink where in several cases the link between a mobile user and a terrestrially placed relay can be weak due to obstacles. In such cases, the user can direct its transmission to a UAV that hovers above a terrestrial relay. Subsequently, the relay amplifies the signal and dispatches it to the destination, where the RIS mounted on the UAV can be utilized again to provide further signal enhancements between the relay and destination.

\subsubsection{Physical layer security (PLS)}

Communications over wireless channels come with the risk of eavesdropping, tampering and forgery. Unlike traditional encryption techniques, PLS offers a low-complexity solution that ensures a secure data transmission even against eavesdroppers with powerful computing tools. 
\par In HRNs, one can think of utilizing the relay as a friendly jammer while two users exchange their data, and the response of RISs can be adjusted to minimize the effect of jamming on the legitimate receiver, while maximizing it at the eavesdropper. Furthermore, in large-scale networks with a large number of relays and RISs, some relays can be selected for relaying operations, while others can act as friendly jammers in a cooperative manner. In such cases, power control among different nodes and RIS configuration can play a key role in the secrecy performance.

\section{Concluding Remarks}
The advantages, challenges, and key applications of different HRNs were introduced and thoroughly discussed in this article. A case study was provided highlighting the rate performance of relay-assisted, RIS-assisted, and HRNs with both AF and DF relaying protocols. Throughout this article, we showed how spatially separated relays and RISs can work harmoniously for enhanced rate performance, and for a variety of different relaying network architectures. Key research directions were provided which can help reveal the true potential of such hybrid networks in future wireless systems.    
	\bibliographystyle{IEEEtran}
	\bibliography{HRNs}
		
\begin{IEEEbiographynophoto}
{Zaid Abdullah} is a Research Associate at the SnT, University of Luxembourg. He obtained his PhD in Communications and Signal Processing from Newcastle University, UK, in 2019. He was a Postdoctoral Research Assistant with the School of Engineering, University of Leicester, UK, between 2019 and 2021. He is passionate about the research on future wireless networks, particularly when it comes to large-scale antenna arrays, integrated terrestrial and non-terrestrial networks, algorithm design, and artificial intelligence. 
\end{IEEEbiographynophoto} 
	
\begin{IEEEbiographynophoto}
{Steven Kisseleff} is a Research Scientist at the SnT, University of Luxembourg. He obtained his Ph.D. degree in electrical engineering from the Friedrich-Alexander University of Erlangen–Nürnberg (FAU), Germany, in 2017. He was a Research and Teaching Assistant with the Institute for Digital Communications, FAU, from 2011 to 2018. In 2012 and 2013, he was a Visiting Researcher with the State University of New York at Buffalo, USA, and the Broadband Wireless Networking Lab, Georgia Institute of Technology, Atlanta, GA, USA. His research activities are focused on the satellite communications, quantum communications, artificial intelligence, and reconfigurable intelligent surfaces. 
\end{IEEEbiographynophoto}
	
\begin{IEEEbiographynophoto}
{Wallace Alves Martins} is a Research Scientist at SnT, University of Luxembourg. He received his D.Sc. degree in electrical engineering from the Federal University of Rio de Janeiro (UFRJ), Brazil, in 2011.  He was an Associate Professor at UFRJ from 2013 to 2022. He serves as an Associate Editor for the Editorial Boards of the IEEE SIGNAL PROCESSING LETTERS and the Journal on Advances in Signal Processing (EURASIP). His research interests include signal processing and telecommunications, with focus on equalization and beamforming/precoding for terrestrial and non-terrestrial networks.
\end{IEEEbiographynophoto}
	
\begin{IEEEbiographynophoto}
{Gaojie Chen} is an Assistant Professor with the Institute for Communication Systems, 5GIC  \& 6GIC, University of Surrey, U.K. He received the Ph.D. degree in electrical and electronic engineering from Loughborough University, UK, in 2012. After his PhD, he took on various academic and research positions at DT Mobile, Loughborough University, University of Surrey, University of Oxford, and the University of Leicester. His research interests include information theory, wireless communications, and satellite communications. He serves as an Associate Editor for the IEEE COMMUNICATIONS LETTERS, IEEE WIRELESS COMMUNICATIONS LETTERS, IEEE JOURNAL ON SELECTED AREAS IN COMMUNICATIONS, and Electronics Letters (IET).
\end{IEEEbiographynophoto}
	
\begin{IEEEbiographynophoto}
{Luca Sanguinetti} is an Associate Professor at the University of Pisa. He was a postdoctoral associate in the Dept. Electrical Engineering at Princeton between 2007 and 2008. From 2013 to 2017, he was with Large Systems and Networks Group (LANEAS), France. He is currently serving as an Associate Editor for the IEEE TRANSACTIONS ON COMMUNICATIONS and is a member of the Executive Editorial Committee of IEEE TRANSACTIONS ON WIRELESS COMMUNICATIONS. His main research interests span the areas of wireless communications and signal processing for communications.  In 2018 and 2022, he received the Marconi Prize Paper Award in Wireless Communications.
\end{IEEEbiographynophoto}
	
\begin{IEEEbiographynophoto}
{Konstantinos Ntontin} is a Research Scientist at the SnT, University of Luxembourg. He obtained his PhD from the Technical University of Catalonia (UPC), Spain, in 2015. In the past, he held Research Associate positions at the University of Barcelona, the University of Athens, and the National Centre of Scientific Research-"Demokritos". In addition, he held an internship position at Ericsson Eurolab Gmbh, Germany. His research interests are related to the physical layer of wireless telecommunications with focus on performance analysis in fading channels, MIMO systems, transceiver design, and stochastic modeling of wireless channels.
\end{IEEEbiographynophoto}
	
\begin{IEEEbiographynophoto}
{Anastasios Papazafeiropoulos} is currently a Vice-Chancellor Fellow with the University of Hertfordshire, UK, and a Visiting Research Fellow with the SnT, University of Luxembourg. He received the Ph.D. degree from the University of Patras, Greece, in 2010. From 2012 to 2014, he was a Research Fellow with Imperial College London, UK, awarded with a Marie Curie fellowship (IEFIAWICOM). He has been involved in several EPSRC and EU FP7 projects such as HIATUS and HARP. His research interests span machine learning for wireless communications, intelligent reflecting surfaces, massive MIMO, heterogeneous networks, among others.
\end{IEEEbiographynophoto}
	
\begin{IEEEbiographynophoto}
{Symeon Chatzinotas} is currently a Full Professor/Chief Scientist I and Head of the SigCom Research Group at SnT, University of Luxembourg. He is coordinating the research activities on communications and networking across a group of 70 researchers, acting as a PI for more than 20 projects and main representative for 3GPP, ETSI, DVB. He is currently serving in the editorial board of the IEEE Transactions on Communications, IEEE Open Journal of Vehicular Technology, and the International Journal of Satellite Communications and Networking. He has (co-)authored more than 600 technical papers in refereed international journals, conferences and scientific books. 
\end{IEEEbiographynophoto}
	
\begin{IEEEbiographynophoto}
{Bj$\ddot{\text{o}}$rn Ottersten} received the Ph.D. degree in electrical engineering from Stanford University, Stanford, CA, USA, in 1990. From 1996 to 1997, he was the Director of Research with ArrayComm, Inc., a start-up in San Jose, CA, USA, based on his patented technology. In 1991, he was appointed Professor of signal processing with the Royal Institute of Technology (KTH), Stockholm, Sweden. Dr. Ottersten has been Head of the Department for Signals, Sensors, and Systems, KTH, and Dean of the School of Electrical Engineering, KTH. He is currently the Director for the Interdisciplinary Centre for Security, Reliability and Trust, University of Luxembourg.
\end{IEEEbiographynophoto}
\end{document}